\documentclass[aps,twocolumn,raggedbottom,prb,amsmath,showpacs,nobalancelastpage,amssymb,groupedaddress]{revtex4}
\usepackage{epsfig}
\begin{document}
\title{Macroscopic quantum effects in a strongly driven %suspended
nanomechanical resonator}
\author{V.\ Peano}
\author{M.\ Thorwart}
\affiliation{Institut f\"ur Theoretische Physik IV, Heinrich-Heine-Universit\"at 
D\"usseldorf, Universit\"atsstra{\ss}e 1, D-40225 D\"usseldorf, Germany}

\date{\today}

\begin{abstract}
We investigate the nonlinear response of a vibrating suspended nanomechanical 
beam on external periodic driving. The amplitude of the 
fundamental transverse mode behaves thereby like a weakly damped 
quantum particle in a driven anharmonic potential. 
Upon using a Born-Markovian master equation, we 
calculate the fundamental mode amplitude for varying driving frequencies. 
In the nonlinear regime, 
we observe resonances which are absent in the 
corresponding classical model. They are shown to be associated with resonant 
 multi-phonon excitations. 
 Furthermore, we identify resonant tunneling in a dynamically induced 
 bistable effective potential.
\end{abstract}
\maketitle
%
%%%%%%%%%%%%% Introduction %%%%%%%%%%%%%%%%%%%%%%%%%%%%%
\section{Introduction}
The ongoing progress in miniaturization of microscale devices 
 allows nowadays to fabricate mechanical resonators on the
nanometer scale \cite{Roukes01}. 
To realize them in form of transversely vibrating 
beams, lithographically patterned doubly clamped suspended 
beams \cite{Roukes00,Kroemmer00,Erbe00,Beil00,Cleland01,Nguyen99,Ming03,
Knobel03,Husain03,LaHaye04} 
are designed, but also suspended %doubly clamped 
carbon
nanotubes display mechanical vibrations \cite{Babic04} 
(see also Ref.\  \onlinecite{Li04}). 
Beyond applications as electrometers \cite{Roukes00,Kroemmer00},  
for detecting ultrasmall forces and
displacements \cite{Beil00,LaHaye04},  or radio-frequency signal
processing \cite{Nguyen99}, the nanomechanical devices  also allow to investigate 
fundamental physical phenomena. In particular, due to their small size, 
the crossover from the classical to the quantum 
regime %of classical physics to the regime where 
%quantum fluctuations in the transverse direction 
%may drastically influence the dynamics 
is of interest  \cite{Carr01,Werner04}. 
The quantum  behavior 
  will be due to a macroscopic number of particles whose 
coherence is disturbed by the interaction with the
environment. % causing damping and decoherence. 
Promising progress in approaching the quantum regime experimentally has been 
reported recently \cite{Ming03,Knobel03,LaHaye04}. 

In this work, we propose 
a straight-forward way to reveal the quantum behavior of nanoresonators 
via the nonlinear response 
of a nanobeam to an external {\em ac\/}-driving. 
 Mechanical excitation frequencies of nanodevices 
 have been measured via {\em ac\/}-driving 
\cite{Kroemmer00,Erbe00,Poncharal99,Beil00,Cleland01,Nguyen99,Ming03,
Knobel03,Husain03,LaHaye04}. 
 We consider 
 the amplitude of the fundamental transverse vibrating mode of 
a doubly clamped suspended nanobeam under longitudinal compression. 
The compression allows to control the degree of nonlinearity of 
the nanoresonator. 
Starting from a continuum model 
\cite{Carr01,Werner04} (see Ref.\  \onlinecite{Nayfeh} for the classical
counterpart), the interacting field theory closed to the 
Euler buckling instability 
can be reduced to %an equivalent 
the dynamics of a quantum particle  
in an anharmonic potential. By exciting the beam to transverse vibrations, 
 its non-linear response can be determined. The 
%influence of the 
environment is included by coupling the system to a
 harmonic Ohmic bath \cite{Weiss99}. 
For weak driving, 
the response will be maximal at the
eigenfrequency of the nanobeam and similar to that
of the analogous  classical system. 
However, as we show below,
for strong driving, the non-linear response of the 
quantum system is qualitatively different from its classical counterpart 
which is the standard Duffing system \cite{Nayfeh}. This feature 
allows to separate both the classical and the quantum regime.
In particular, we find distinct resonances in the  dependence of
the  amplitude of the fundamental mode on the driving frequency. 
They can be interpreted as resonant multi-phonon excitations 
and are determined by avoided level crossings 
in the Floquet spectrum. 
In turn, we identify a separation of time scales and 
resonant tunneling in the dynamically 
induced bistable effective potential.  
%%%%%%%%%%%%%%%%%%%%%%%%%%%%%%%%%%%%%%%%%%%%%%%%%%%%%%%%

\section{Effective single-particle model starting from a continuous model}

We start from the normal mode description of an elastic rectangular beam 
of length $l$, width $w$ and thickness $d$ such that $l \gg w>d$ 
\cite{Carr01,Werner04,Nayfeh}, which 
 permits to consider only transverse displacements of the beam. 
A longitudinal mechanical force 
controls the nonlinearity of the potential energy in the transverse 
direction. At low temperatures, the higher modes are frozen out and the fundamental
mode can be treated independently. The resulting effective potential
energy of the fundamental mode contains terms which are quadratic  and 
quartic in the transverse amplitude $x$ of the  fundamental mode. 
The externally applied {\em ac\/}-driving can be included on the same footing. 
%: 
%for instance, an {\em ac\/}-current can be  applied in the longitudinal direction  
%together with a constant magnetic field in the $z$-direction. The
% Lorentz force then induces transverse displacements. 
 The driving strength can be tuned within the fundamental mode description to the regime where
non-linear effects come into play but the higher modes are still
negligible (see below). 
One ends up with an effective single particle quantum mechanical
time-dependent Hamiltonian for the fundamental mode amplitude $x$       
acting as a position operator, i.e., \cite{Carr01,Werner04}
\begin{equation}
H_S(t)=-\frac{\hbar^2}{2m^*}\frac{\partial ^2}{\partial x^2}
+\frac{\tilde{\alpha}}{2}x^2+\frac{\tilde{\beta}}{4}x^4+ x f \cos(\omega
t) \, . \label{ham}
\end{equation}
Here, $m^*$ is the effective mass of the beam. 
The parameters in the
potential are obtained as \cite{Carr01,Werner04}
$\tilde{\alpha}=m^*\omega_0^2 \equiv m^* \overline{\omega}_0^2 
\left(\frac{\varepsilon_c-\varepsilon}{\varepsilon_c}\right) %\; \;
$ %
and $
\tilde{\beta} = \frac{3 m^* 
\overline{\omega}_0^2}{d^2}, \, 
$%
where the frequency of the 
fundamental mode is given by $\overline{\omega}_0=\pi^2 d 
\sqrt{{\cal Q}/\rho}/ (l^2 \sqrt{12})$. ${\cal Q}$ is Young's elasticity modulus and $\rho$ the 
mass density of the material. 
The longitudinal force generates the strain $\varepsilon=(l-l_0)/l_0$ where $l_0$ is the 
equilibrium length of the beam. At the critical value  $\varepsilon_c=-\pi^2 d^2 / (12 l^2)$ 
(for a rectangular beam), the system reaches a bifurcation point which is the well-known Euler 
instability. The effective potential for the fundamental mode then is purely quartic. 
Close to the Euler instability, both quadratic and quartic terms appear. 
We consider the case of a monostable potential, i.e., $\alpha > 0$, which 
implies that the strain remains below its critical values, i.e., $\varepsilon
<\varepsilon_c$. This situation is easier to be realized experimentally compared to the 
bistable potential $\alpha < 0$. The {\em ac\/}-driving occurs with amplitude 
$f$ and frequency $\omega$. An upper limit $f_{\rm max}$ for the regime of validity of the 
fundamental mode description is given by the first harmonic threshold, i.e., 
$x_0 f_{\rm max} < \hbar \omega_1 \approx 3 \hbar \omega_0$ \cite{Carr01}. 
We scale the Hamiltonian (\ref{ham}) with respect to the units 
in space and  time, i.e., $x_0=\sqrt{\hbar/(m^* \omega_0)}$ and $t_0=1/\omega_0$, 
 respectively, yielding the energy scale $\hbar \omega_0$ and the dimensionless nonlinearity 
 parameter 
$ \beta\equiv\frac{3}{4 (d/x_0)^2} \frac{\varepsilon_c}{\varepsilon_c-\varepsilon}$. 
%

%In a realistic model  also the effect of the  
 %environment has to be included. We describe this  phenomenologically 
 We include the effect of the environment phenomenologically 
by a set of harmonic oscillators which are bilinearly coupled to the system  
 with the coupling constants $c_j$ 
\cite{Weiss99}. %This picture arises from the normal mode description of the beam where 
%all the higher modes 
%
The  Hamiltonian is
\begin{equation}
H_B=\frac{1}{2}\sum_j p_j^2/m_j + m_j \omega_j^2 \left( x_j- \frac{c_j}{m_j \omega_j^2}x\right)^2. %\, .
\end{equation}
The bath is characterized by  the 
spectral density 
\begin{equation}
I(\omega)=\frac{\pi}{2}\sum_j\frac{c_j^2}{m_j\omega_j}\delta(\omega-\omega_j)=
m^*\gamma\omega e^{-\omega/\omega_c}\, , 
\end{equation}
where we have chosen the standard Ohmic 
form with damping constant $\gamma$ and  with  cut-off 
frequency $\omega_c$.  The total Hamiltonian is $H(t)=H_S(t)+H_B$. 

We focus on the 
case 
when the coupling to the 
bath is weak and use a standard Born-Markovian master equation 
\cite{Louisell,Kohler97} which reads
\begin{equation}
\dot{\rho}=-\frac{i}{\hbar}\left[ H_S(t),\rho \right]+ 
 {\cal{L}}_{\rm rel} (\rho)+{\cal{L}}_{\rm noise}(\rho) \, . 
\end{equation}
The commutator describes the coherent 
dynamics while  the bath acts via the 
two superoperators for relaxation and noise, respectively, 
\begin{eqnarray}
{\cal{L}}_{\rm rel} (\rho) &=& -\frac{i \gamma}{2 \hbar} \left[ x,
\left[p,\rho\right]_+ \right] \, , \nonumber \\
{\cal{L}}_{\rm noise} (\rho) & = & -\frac{1}{\hbar} \left[ x, \left[Q,\rho\right]
\right] \, .
\end{eqnarray}
The operator %
\begin{equation}
Q=\int_0^\infty d\tau K(\tau) x_{\rm H}(t-\tau,t)
\end{equation}
 involves the 
integral kernel 
\begin{equation}
K(\tau)=\frac{1}{\pi}\int_0^\infty d\omega I(\omega) \coth 
\left( \frac{\hbar \omega}{2 k_B T}\right) \cos \omega t
\end{equation}
with $T$ being temperature 
and the position operator $x_{\rm H}(t-\tau,t)$ in the Heisenberg picture. 
This master equation can be solved numerically to obtain  
$\rho(t)$ and 
%Then, 
%we can evaluate 
the expectation value 
\begin{equation}
\langle x(t)\rangle=tr (\rho(t) x)\, .
\end{equation}
%  
%of the position operator. 
In the stationary limit,  $\langle x(t)\rangle$ shows oscillations with 
the frequency being the external driving frequency $\omega$ (plus a phase shift) 
and with amplitude $A$. This amplitude is 
the quantity of interest that allows to compare 
to the classical counterpart. 
\begin{figure}[t]
\begin{center}
\epsfig{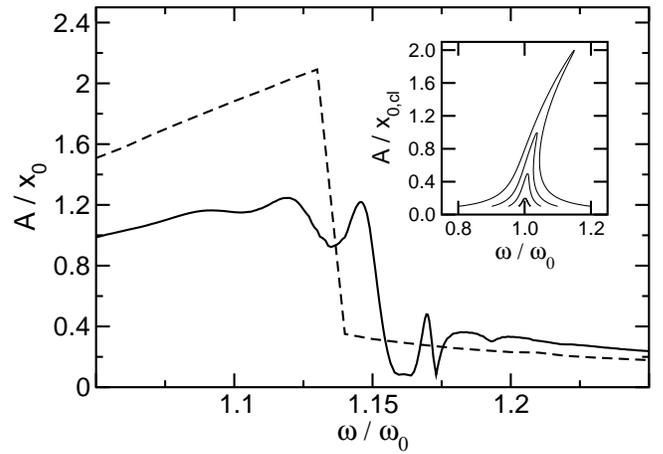}
\caption{Amplitude $A$ of the expectation value $\langle x (t)\rangle$ 
%in units of $x_0=\sqrt{\frac{\hbar}{m^*\omega_0}}$ 
for varying the driving frequency $\omega$. 
Parameters are $k_B T=0.1 \hbar \omega_0, 
\beta=0.1, f=0.1 \hbar \omega_0 / x_0, \gamma = 0.005 \omega_0$. 
Dashed line: Results of the classical Duffing oscillator at 
$T=0$ with the remaining parameters being the same. 
%for the same parameters. 
Inset: Amplitude $A$ of the classical Duffing oscillator for varying 
driving frequencies. The driving strength $f$ increases from bottom 
to top. \label{fig.1}}
\end{center}
\end{figure}

%It is well known that 
\section{Non-linear response of the nanoresonator}
With the oscillation amplitude $A$ at hand, we can investigate the non-linear
response of the nanoresonator on varying driving frequencies $\omega$. 
The classical analogue 
shows a well-known rich dynamics including chaos and 
driving induced bistability. 
 Without noise, we obtain the 
Duffing oscillator \cite{Holmes,Nayfeh}. Its response on varying the driving 
frequency can be calculated perturbatively   \cite{Nayfeh} 
and is shown in  Fig.\ \ref{fig.1} (dashed line and inset). 
For weak driving, the standard Lorentzian 
resonance of a harmonic oscillator occurs. For stronger driving,  
the system enters the nonlinear regime, the resonance curve bends over and an effective 
bistability is induced  yielding a 
 hysteretical response. If 
Gaussian white noise is added, the range of bistability and therefore hysteresis is reduced 
\cite{Datta01} since the system escapes more easily from the metastable state 
% bistability is more easily 
%overcome 
by thermal hopping. 

For the quantum case, one can ask whether a signature of the 
induced bistability still appears in the response.  
For a truly bistable potential, tunneling 
 leads to a reduction of  hysteresis \cite{Thorwart97}.  
Also, resonant tunneling through the potential barrier leads 
to characteristic steps in the hysteresis cycle \cite{Thorwart98}. 
%The quantum steps  
%have  been experimentally detected and  associated to  
% Macroscopic Quantum Tunneling of the magnetization 
%of organic high-spin molecular crystals \cite{Spin}.
However, the potential in the Hamiltonian Eq.\  (\ref{ham}) is monostable. 

\begin{figure}[t]
\begin{center}
\epsfig{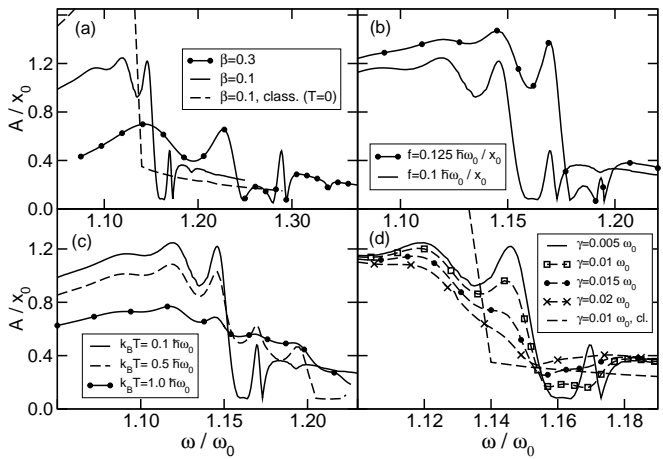}
\caption{Amplitude $A$ %of the expectation value $\langle x (t)\rangle$ 
for varying the driving frequency $\omega$ and for different 
choices of (a) the nonlinearity coefficient $\beta$, (b) driving 
intensity $f$, (c) temperature $T$ and (d) damping strength $\gamma$. The other 
parameters are kept fixed according to $\beta=0.1, f=0.1 
\hbar \omega_0 / x_0, k_B T =  0.1 \hbar \omega_0$ 
and $\gamma=0.005 \omega_0$, respectively. Dashed line in (a) and (d): 
same as in Fig.\  \ref{fig.1}. 
 \label{fig.2}}
\end{center}
\end{figure}
We have calculated the amplitude $A$ of the steady-state oscillations 
 for varying driving frequencies $\omega$. 
The result is shown in Fig.\  \ref{fig.1}. 
The characteristic profile consists of a series of peaks and dips. 
The resonances are sharper for higher frequencies. For lower frequencies, 
the broad peaks overlap strongly and lead to a shoulder-like profile which 
 is similar to the classical result (dashed line in Fig.\  \ref{fig.1}). 
 The locations of the resonances are dominated by the parameters of the 
undamped driven system, see 
%. This can be seen in 
Fig.\  \ref{fig.2}a and \ref{fig.2}b. %, where the 
%nonlinearity coefficient $\beta$ 
%and driving intensity $f$ have been changed, respectively. 
%
On the other hand, by changing the parameters 
characterizing the bath (temperature $T$ and damping strength $\gamma$)
 the shape of the resonances is modified.  
 As can be seen in Fig.\  \ref{fig.2}c and  \ref{fig.2}d, for  increasing 
  temperature and damping, the sharp resonances are smeared out 
and finally fade out when the quantum coherence is suppressed. 
 This indicates that the 
  resonances are related to resonant multi-phonon excitations in the 
  driven nonlinear system. 

\begin{figure}[t]
\begin{center}
\epsfig{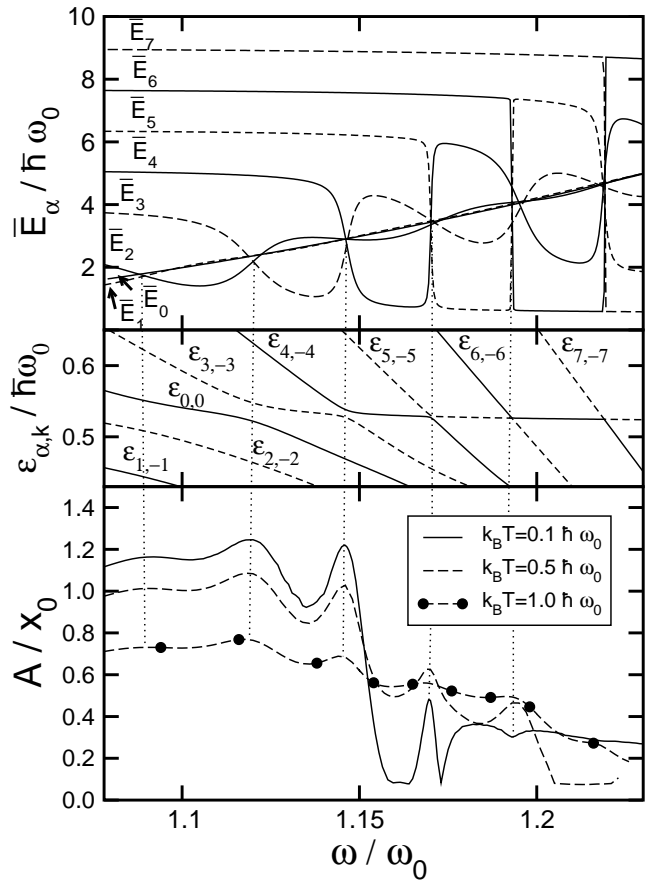}
\caption{Average energies  $\overline{E}_i$ (top), 
quasienergy levels $\varepsilon_{\alpha, k}$ (middle) and 
amplitude $A$ of the fundamental mode for 
 varying driving frequencies $\omega$.  
The remaining parameters are $\beta=0.1, f=0.1 \hbar \omega_0 / x_0$ 
and $\gamma=0.005 \omega_0$. \label{fig.3}}
\end{center}
\end{figure}

\section{Resonant multi-phonon excitation}
In order to show this, we 
exploit the periodicity in time of %the 
%system Hamiltonian 
$H_S(t)$ and calculate the quasienergy (Floquet) spectrum 
for varying driving frequencies $\omega$ \cite{Floquet,Fainshtein78}. To this 
end, we solve the equation 
\begin{equation}
[ H_S(t) -i\hbar \frac{\partial}{\partial t}]
|\phi_\alpha (t) \rangle = \varepsilon_\alpha |\phi_\alpha (t) \rangle \, ,
\end{equation}
where the quasienergies $\varepsilon_\alpha$ are defined up to a multiple integer 
of $\hbar \omega$. %the energy quantum of the {\em ac\/}-field. 
This means that the state $
|\phi^{(n)}_{\alpha}(t)\rangle 
= e^{i n \omega t}|\phi_{\alpha}(t)\rangle
$ 
is also an eigenstate of the Floquet Hamiltonian, but with the eigenvalue 
\begin{equation}
\varepsilon_{\alpha,n}=\varepsilon_\alpha + n \hbar \omega \, .
\end{equation}
Since the quasienergies do not allow for 
global ordering, we have also calculated the mean energies 
averaged over one driving period 
\begin{equation}
\overline{E}_\alpha
=\sum_n (\varepsilon_{\alpha}+n \hbar \omega) \langle c_{\alpha,n} | 
c_{\alpha,n}\rangle\, ,
\end{equation}
where the $|  c_{\alpha,n}\rangle$ are the Fourier components of the Floquet states 
\cite{Floquet}, which can be obtained as
\begin{equation}
|  c_{\alpha,n}\rangle = \frac{\omega}{2 \pi} \int_0^{2\pi/\omega} dt e^{in
\omega t } |\phi_{\alpha}(t)\rangle \, .
\end{equation}
%. 
For the harmonic case ($\beta=0$), the quasienergies are all 
degenerate for infinitesimal driving. The strong driving 
and the nonlinear potential lift this degeneracy.  
As follows from the result shown in Fig.\ \ref{fig.3},  
the quasienergies show avoided level crossings 
(and the mean energies show exact crossings) 
for particular values of the driving frequency $\omega$. 
To these (anti-)crossings correspond the particular resonances in the 
frequency-dependence of the amplitude $A$. 
The resonances occur when the (undamped) quantum system absorbs 
an integer multiple of %the energy quantum 
$\hbar \omega$. 
The shape of the resonances is related to the splitting of the 
quasienergy levels at the avoided level crossings. For lower 
frequencies, the level splitting is quite large. Moreover, 
the avoided crossings are not well separated in this regime. 
This implies that the resonances are broad and strongly overlap yielding 
 the shoulder-like behavior of the response profile. % which  is similar to the 
%classical result
%, see Fig.\  \ref{fig.1}. 
For 
%higher driving frequencies
larger $\omega$, the avoided level crossings become 
well separated and the energy splittings decrease. 
Note that the resonances occur despite the fact that the thermal energy  
is larger than the quasienergy level splittings at the avoided level crossing. 
In turn, this leads 
to sharp resonances in the amplitude. The physics of avoided 
quasi-energy level crossings has been discussed in detail in Ref.\  
\onlinecite{Kohler98}. 

\begin{figure}[t]
\begin{center}
\epsfig{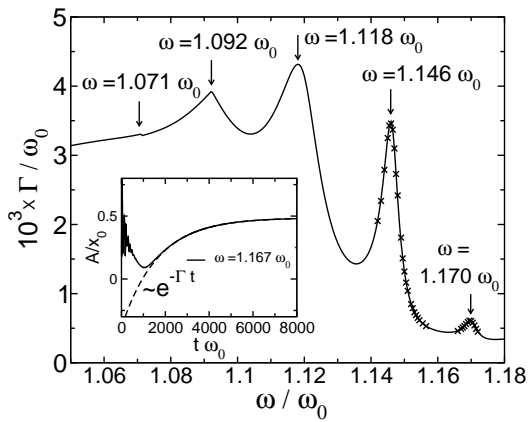}
\caption{Decay rate  $\Gamma$ for the slow dynamics of the 
amplitude $A$ for approaching the steady state. Symbols: 
Solution from the numerical iteration of the Born Markovian 
master equation, solid line: Smallest non-zero eigenvalue  
of the rate matrix of an improved Floquet Markovian master equation 
(see text). 
Inset: time-resolved dynamics of $A$ for $\omega=1.167 \omega_0$ 
(solid line). The dashed line shows a fit to an exponential 
$~e^{-\Gamma t}$. Here, 
$\beta=0.1, f=0.1 \hbar \omega_0 / x_0, k_B T=0.1 \hbar \omega_0$  
and $\gamma=0.005 \omega_0$. \label{fig.4}}
\end{center}
\end{figure}
%
%%%%%%%%%%%%%%%%%%%%%%%%%%%%%%%%%%%
\section{Dynamically induced bistability and Resonant Tunneling}
The bistability of the steady state of the classical 
Duffing oscillator does not survive in the quantum system 
since it escapes the metastable state asymptotically 
via tunneling \cite{Thorwart98}. 
Nevertheless, we find signatures of  bistability 
and tunneling if we consider how the steady state is reached. 
For this, we show in the inset of Fig.\  \ref{fig.4} 
the amplitude $A$ (local maxima of the vibrations) 
for increasing time (starting with the ground state of the 
undriven oscillator as the initial state). We observe fast 
oscillations at short times. They decay on a time scale 
$\gamma^{-1}$ which reflects intrawell relaxation 
in the metastable state. 
 Then, starting from a metastable 
state at intermediate times, a slow exponential 
decay towards the asymptotically stable state can be observed.  
By fitting to an exponential, we extract the decay rate $\Gamma$ for 
various driving frequencies, see Fig.\  \ref{fig.4} crosses. 
The decay rate shows resonances at the same values of the frequencies 
where the avoided crossings of the quasienergy levels occur 
(see arrows). 
This slow dynamics can be identified as quantum tunneling 
in a dynamically induced bistable effective potential 
(see also Ref.\  \onlinecite{Rigo97}). % for a related feature in a 
%different nonlinear oscillator). 
The 
peaks in $\Gamma$ indicate resonant tunneling 
\cite{Thorwart98} 
from the meta- to the globally stable state 
both of which are dynamically induced.

The rate for tunneling out of the metastable well can be obtained 
 directly from an improved Floquet-Markovian master equation 
\cite{Kohler97}. In the moderate rotating wave approximation, the  
time-independent master equation 
\begin{equation}
\dot{\rho}_{\alpha\beta} (t) = -\frac{i}{\hbar} 
\left(
\varepsilon_\alpha - \varepsilon_\beta
\right) \rho_{\alpha\beta} (t) + 
\sum_{\alpha' \beta'} 
{\cal L}_{\alpha\beta,\alpha' \beta'} \rho_{\alpha' \beta'} (t) \, 
\label{floqme}
\end{equation}
in the Floquet basis can be established. The transition rates containing 
the influence of the dissipative bath read\cite{Kohler97}
\begin{eqnarray}
{\cal L}_{\alpha\beta,\alpha' \beta'} & = &  \sum_n \left( N_{\alpha \alpha', n} +
N_{\beta \beta', n} \right) X_{\alpha \alpha', n} X_{\beta' \beta, -n} \nonumber \\
& & - \delta_{\beta \beta'} 
\sum_{\beta'', n}  N_{\beta'' \alpha', n}  X_{\alpha \beta'', -n} X_{\beta''
\alpha', n} \nonumber \\
& & - \delta_{\alpha \alpha'} 
\sum_{\alpha'', n}  N_{\alpha'' \beta', n}  X_{ \beta' \alpha'', -n} X_{\alpha''
\beta, n} \, . 
\label{ratecoeff}
\end{eqnarray}
Here, the coefficients are given by
\begin{eqnarray}
N_{\alpha \beta, n} &=& N(\varepsilon_\alpha-\varepsilon_\beta + n \hbar \omega)\,
, \mbox{\hspace{1ex}} N(\varepsilon) = \frac{m^* \gamma \varepsilon}{\hbar^2} 
\frac{1}{e^{\varepsilon/k_B T}-1} \,  \nonumber \\
X_{\alpha \beta, n} &=& \frac{\omega}{2 \pi} 
\int_0^{2\pi/\omega} dt e^{-in\omega t} \langle \phi_{\alpha}(t)|x|\phi_{\beta}(t)\rangle
\, .
\end{eqnarray}
The time-independent rate coefficients (\ref{ratecoeff}) 
can be written as a matrix which can readily be diagonalized numerically. We
find a clear separation of time-scales where the smallest non-zero eigenvalue 
indicates a slow tunneling dynamics by which the stationary state is approached.
% which has constant coefficients. 
The solid line of Fig.\  \ref{fig.4} shows the smallest non-zero eigenvalue 
of the corresponding rate matrix which was obtained by 
direct numerical diagonalization using a basis of 8 Floquet eigenstates. 
The result agrees well with that from the numerical iteration of 
the Born-Markovian master equation. Note again the analogy to resonant tunneling
in a static double-well potential \cite{Thorwart98}. The role of the
eigenenergies in the static case is now played by the quasienergies
$\varepsilon_\alpha$ determining the coherent dynamics, see Eq.\  
(\ref{floqme}). In both cases, the avoided (quasi-)energy level crossings are
the origin of resonant tunneling from the metastable towards the globally stable
state. Nevertheless, the incoherent part of Eq.\  (\ref{floqme}) is crucial to
observe the resonant tunneling in this driving induced bistability. 

\section{Conclusions}
In conclusion, we have found resonances in the amplitude of the fundamental mode of 
a nanomechanical resonator which is driven into its non-linear regime. They occur for 
particular values of the driving frequency and  
are explained in terms of quantum mechanical resonant multi-phonon excitations between 
quasienergy states.  This response profile is generic for the quantum system 
and is absent in the  analogous classical system. 
 In addition, we have identified resonant tunneling in a dynamically induced 
 bistable effective potential.
Since oscillation amplitudes of nanomechanical 
 resonators can be detected with currently available experimental techniques, 
 this macroscopic quantum effect should be measurable in the 
 near future. A promising approach seems to be to couple a superconducting 
  single-electron transistor capacitively to the resonator \cite{LaHaye04} 
  allowing an ultra-high sensitivity. Indeed, 
  a fundamental frequency of 19 MHz has been reported \cite{LaHaye04} 
  implying that quantum effects occur below a temperature of a few mK. 
  Most interestingly, the reported quality factors around $Q=40000$ seem to be
  very promising since we predict quantum effects at even smaller 
  $Q=\omega_0/\gamma$, see above.   \\
  Finally, 
  we note that our model of a driven anharmonic mono-stable quantum oscillator
  is a generic model which finds applications in various 
  other fields of physics. One important  application is related to 
  the non-destructive readout for superconducting flux or charge qubits
  \cite{Lupascu03}. There, a dc-SQUID which is inductively coupled to the
  qubit is driven by an ac bias current such that the SQUID remains in
  its superconducting state (for this, the amplitude of the ac current has to be
  smaller than the critical current of the SQUID). 
  The Josephson inductance of the SQUID carrying information on
  the qubit state is measured for varying the frequency of the ac-current. 
  In terms of our model, the SQUID provides a sinusoidal potential for the 
  superconducting phase acting as a single quantum mechanical particle. 
  The particle is initially localized in one potential minimum. The
  ac bias current provides the external driving and for
  stronger driving, the  particle experiences the nonlinearity of the potential
  well. The distinct resonances in the nonlinear response which we have found
  might help to increase the efficiency of the read-out process since they are
  sharper than the common linear resonance.   \\
\begin{acknowledgments}
\vspace*{-5mm}
We thank Peter H\"anggi for helpful discussions. 
We  acknowledge support %by the Deutsche Forschungsgemeinschaft (M.T., Th820/1-1)  
%and 
by the DFG-SFB/TR 12.% (V.P.).
\end{acknowledgments}
%

%%%%%%%%%%%%%%%%%%%%%%%%%%%%%%%%%%%%%%%%%%%%%%%%%%
\end{document}